\documentstyle[12pt]{article}
\renewcommand{\section}[1]{\refstepcounter{section}
\vspace{24pt}\noindent{\bf\arabic{section}.\quad #1}
\vspace*{12pt}}
\newcommand{\ulsect}[1]{\vspace{18pt}\noindent{\bf #1}
\vspace*{12pt}}

\begin{document}
\begin{flushright} CERN-TH.6882/93\\
\end{flushright}
\vspace*{10mm}
\begin{center}
{\bf Thermal quark production in pure glue and quark gluon
plasmas}\\[10mm]
Tanguy Altherr$^*$ and David Seibert$^{\dag}$\\[5mm] Theory
Division, CERN, CH-1211 Geneva 23, Switzerland\\[10mm]
{\bf Abstract}\\
\end{center}
\hspace*{12pt}We calculate production rates for massless $(u,d)$
and massive $(s,c,b)$ quarks in pure glue and quark gluon plasmas
to leading order in the strong coupling constant $g$.  The
leading contribution comes from gluon decay into $q\bar q$ pairs,
using a thermal gluon propagator with finite thermal mass and damping
rate. The rate behaves as $\alpha_S^2(\ln 1/\alpha_S)^2 T^4$ when $m,
\alpha_S \rightarrow 0$ and depends linearly on the transverse gluon damping
rate for all values of the quark mass $m$.  The light quark ($u$, $d$, $s$)
chemical equilibration time is approximately 10-100 $T^{-1}$ for $g=$2-3,
so that quarks are likely to remain far from chemical equilibrium in
 ultrarelativistic nuclear collisions.\\
\vfill
\begin{center}
{\em Submitted to Physics Letters B}
\end{center}
\vfill
CERN-TH.6882/93\\
April 1993\\
\vspace*{10mm}
\footnoterule
\vspace*{3pt}
\noindent$^*$On leave of absence from  L.A.P.P., BP110, F-74941
Annecy-le-Vieux Cedex, France. Internet: taltherr@vxcern.cern.ch.\\
$^{\dag}$On leave until October 12, 1993 from:
Physics Department, Kent State University, Kent, OH 44242 USA.
Internet: seibert@surya11.cern.ch.\\
\newpage\setcounter{page}{1} \pagestyle{plain}
     \setlength{\parindent}{12pt}


Ultrarelativistic nuclear collisions may present a unique opportunity
to study the high temperature behaviour of hadronic matter in the
laboratory.  It is possible that in these experiments, the produced
energy densities might be high enough and last for long enough for
bulk, locally equilibrated, deconfined matter, or quark gluon plasma
(QGP), to be created and studied.  However, recent work
[\ref{rg},\ref{rs}] has shown that, while the hot matter should
thermalize in about 0.3 fm/$c$, it may be largely a gluon plasma (GP),
with few quarks.  Chemical equilibration is expected to take much
longer than thermal equilibration, if it occurs at all.

One of the proposed signals for QGP is copious production of strange
particles [\ref{rstrange}], as the strangeness production rate is
much higher in QGP than in normal hadronic matter.  Similarly, it
has also been argued that thermal charm production in a QGP could
be large [\ref{rcharm}].  However, these rates have never been
calculated using thermal field theory.

In this paper, we calculate the thermal quark production rate to
leading order in $\alpha_S$ for GP and for QGP.
This gives the chemical equilibration rates for both the massless
($u$, $d$) and massive ($s$, $c$, $b$) quark components of QGP.  We
find that these rates are small; the light quarks ($u$, $d$, $s$)
chemical equilibration time is approximately 10-100 $T^{-1}$ for
$g=$2-3, so that quarks are likely to remain far from chemical
equilibrium in ultrarelativistic nuclear collisions.

We take the idealized situation of a GP or QGP in thermal
equilibrium.  If such a plasma does exist, the massless gluons evolve
into quasi-particles with effective masses of order $gT$ [\ref{Plasmon}].
Being massive, these quasi-gluons decay into $q\bar q$ pairs. This
situation is similar to the plasmon decay into $\nu\bar\nu$ pairs that is
the dominant cooling mechanism for high density stars [\ref{Bra}].

For gauge field theories at high temperature, there exists a resummation
program developed by Braaten and Pisarski [\ref{BP}] that describes
these quasi-particles.  It is easy to check that this resummation program
is almost the same for the cases of QGP and GP.  The difference is that the
quark contribution inside each hard thermal loop is omitted in the case of
a GP, because there are no thermal quarks. For instance, the plasma
frequency is
\begin{equation}
       \omega_0^2 = \left( N+ {N_f\over 2} \right) {g^2 T^2\over 9},
\end{equation}
for $SU(N)$ gauge theory at temperature $T$, where $g$ is the strong
coupling constant and $N_f$ is the number of massless fermion flavours.
Thermal effects in QGP and GP are thus identical, except that $N_f=0$ for
GP.

The discussion of the different QCD corrections to the production of
$q\bar q$ pairs is best illustrated by using ``cut'' diagrams
[\ref{Wel},\ref{KS}].  The lowest order diagram is shown in Fig.~1.  To
disentangle the contributions from different graphs and from different
discontinuities within the same graph, we consider only the leading term
in the strong coupling constant $g$ as $g \rightarrow 0$.

Following the usual treatment of Braaten and Pisarski [\ref{BP}] in such
a situation, we consider the two cases when the gluon momentum is soft,
$Q\sim gT$, and hard, $Q\sim T$. If it is soft, one has to use the
effective quark-gluon vertices and gluon propagators which resum the
hard thermal loops [\ref{BP}]. One is therefore left with a quite
involved calculation as several discontinuities can be taken in such
a case. In principle, this situation should be very similar to previous
calculations such as photon emission from a QGP [\ref{Photon}]. Indeed,
the na\"\i ve lowest order diagrams for the $q\bar q$ production are the
two-gluon fusion processes shown in Fig.~2. However, it is well known
that these diagrams have a logarithmic singularity when the quark mass,
$m$, goes to zero [\ref{Com}]. The cross section for this process is
\begin{equation}
\sigma_{gg\to q\bar q} = {\pi\alpha_S^2\over 3s} \ln{s\over m^2}
         \qquad \mbox{when} \quad s\gg m^2
.\end{equation}
The divergence being of kinematic origin, it is clear that in the case of
unthermalized and zero mass quarks, this singularity can be screened only
by the thermal gluon mass, of order $gT$. As typical gluon energies are of
order $T$, the quark production rate must then behave as $g^4 T^4 \ln 1/g$.
When the unthermalized outgoing quarks are massive, the cross section for
the processes in Fig.~2 is no longer singular and leads to a contribution
of ${\cal O}(g^4T^4)$ to the production rate.

A consistent calculation would require a division of the processes into
two pieces by the introduction of an intermediate cutoff $gT\ll q^*\ll T$
on the gluon line (as in the photon emission case). Then the hard part,
calculated from the first two graphs shown in Fig.~2, gives a
contribution of order $g^4 T^4 \ln(T/q^*)$.
On the other hand, the soft part, calculated
from the discontinuity of the graph in Fig.~1 together with effective
vertices and gluon propagator, must give $g^4 T^4 \ln(q^*/\omega_0)$ with
the same coefficient in front so that, by adding the two contributions,
the $q^*$ dependence cancels. To be honest, we have not checked that the
coefficient in front of the logarithm is the same for the soft and the
hard pieces but we do not see any reason why it should differ.

If the gluon momentum is hard and if one is using a bare gluon propagator,
the rate vanishes for obvious kinematic reasons. However, as in the $\gamma
\to \nu\bar\nu$ process, one could also use a resummed propagator for the
hard gluon, leading to a finite answer (in fact a contribution of order
$g^4 T^4$ as we shall see). But unlike the photon, the gluon has an
anomalously large damping rate and if the hard thermal gluon mass starts to
be relevant, one must also include the damping rate inside the hard gluon
propagator. We show that the rate behaves then as $g^4 T^4 (\ln 1/g)^2$
for massless quarks and as $g^4 T^4 \ln 1/g$ for massive quarks.
It is therefore {\it larger} than the soft gluon contribution and than
the na\"\i ve lowest order diagrams shown in Fig.~2. The precise
reasons for this unusual behaviour will become clear as the calculation
proceeds.

The gluon propagator that we use is therefore
\begin{equation}
  {-i{\cal P}^{\mu\nu}_{T(L)} \delta_{ab}\over Q^2 - \mbox{Re} \Pi_{T(L)}
(\omega, q) +2iq^0\gamma_{T(L)}},
\end{equation}
where ${\cal P}^{\mu\nu}_{T(L)}$ is the transverse (longitudinal)
projector [\ref{Plasmon}] and $a$ and $b$ are colour indices.
Here (and for the rest of this paper) the subscript $T(L)$ refers to
transversely (longitudinally) polarized gluons. We have included the
thermal gluon mass at hard momentum, $\mbox{Re}\Pi_{T(L)}
(\omega, q)$, and also the anomalously large gluon damping rate,
$\gamma_{T(L)}$ [\ref{Pis}].

Using the cutting rules of Kobes and Semenoff [\ref{KS}], the quark
(antiquark) production rate due to transverse gluon decay is given by
\begin{eqnarray}
R_T^{\mbox{q}} = \frac {dN_T^{\mbox{q}}} {d^4x} = g^2 \,
\int \frac {d^4Q} {(2\pi)^4} \int \frac {d^4K} {(2\pi)^4} \,
(2\pi)^2 \, \delta [K^2-m^2] \, \delta [(Q-K)^2-m^2] (n_B (\omega)
+\theta(-q^0)) \nonumber \\
\times \mbox{Re} \frac {2i} {Q^2-\mbox{Re}\Pi_T(Q) + 2i q^0
\gamma_T} \, (T^a_{bc})^2 \, \mbox{Tr} \left[ (\not\!\!{K}+m)
\gamma_{\mu} (\not\!\!{K}-\not\!\!{Q}+m) \gamma_{\nu} \right] \,
{\cal P}^{\mu \nu}_T.
\end{eqnarray}
Here $K$ and $Q$ are the quark and gluon four-momenta respectively,
$m$ is the quark mass, and $n_B$ is the Bose-Einstein distribution function.
The transverse gluon projection operator is ${\cal P}^{ij}_T =
-\delta^{ij} + q^i q^j/q^2$, with all other components zero, and
$T^a_{bc}$ are the colour $SU(3)$ matrices.  We use the standard high
energy conventions that $\hbar=c=k_B=1$.

The integrals over $q^0$ and $\cos\theta=\hat{k} \cdot \hat{q}$
can be evaluated by using the two $\delta$-functions, while the
remaining angular integrals are all trivial:
\begin{eqnarray}
R_T^{\mbox{q}} = \frac {g^2} {6\pi^4} \, \int_{2m}^{\infty}
d\omega \, n_B (\omega) \, \int_0^{\sqrt{\omega^2-4m^2}}
dq \, q \, \frac {\omega \gamma_T}
{\left[ \omega^2 -q^2 -\mbox{Re}\Pi_T(Q) \right]^2
+4 \omega^2 \gamma_T^2} \nonumber \\
\times \int dE \, \mbox{Tr} \left[ (\not\!\!{K}+m) \gamma_{\mu}
 (\not\!\!{K}-\not\!\!{Q}+m) \gamma_{\nu} \right] \,
{\cal P}^{\mu \nu}_T.
\end{eqnarray}
The trace over colour indices gives a factor 4/3, the quark energy
$k^0=E=(k^2+m^2)^{1/2}$, and the gluon energy $q^0=\omega$.
The transverse component of the trace is
\begin{eqnarray}
\mbox{Tr} \left[ (\not\!\!{K}+m)  \gamma_{\mu}
(\not\!\!{K}-\not\!\!{Q}+m) \gamma_{\nu} \right] \,
{\cal P}^{\mu \nu}_T = 4 \left[ 2K_{\mu}K_{\nu} - K_{\mu}Q_{\nu}
- Q_{\mu}K_{\nu} + K \cdot Q g_{\mu \nu} \right] \,
{\cal P}^{\mu \nu}_T \nonumber \\
=4 \left[ \omega^2 -q^2  -2k^2 (1-\cos^2\theta) \right],
\end{eqnarray}
where
\begin{equation}
\cos\theta ~=~ (2\omega E - \omega^2 + q^2)/(2kq).
\end{equation}
We then integrate the trace over the allowed range of $E$,
where $\left| \omega^2 -q^2 -2E\omega \right| \leq 2kq$:
\begin{equation}
R_T^{\mbox{q}} = \frac {4g^2} {9\pi^4} \int_{2m}^{\infty}
d\omega\ \omega n_B (\omega) \int_0^{\sqrt{\omega^2-4m^2}}
dq\ q^2 \frac { \gamma_T \left( \omega^2-q^2+2m^2 \right)
\left[ 1- \frac {4m^2} {\omega^2-q^2} \right]^{1/2}}
{\left[ \omega^2 -q^2 -\mbox{Re}\Pi_T(Q) \right]^2
+4 \omega^2 \gamma_T^2}. \label{elm}
\end{equation}

The integral in eq.~(\ref{elm}) is cut off by $n_B$ when $\omega/T \gg 1$.
The crucial point of our analysis is that the leading contribution comes
when $\omega$ and $q$ are hard, of order $T$. The soft $Q$ contribution is
perfectly regular and subleading by ${\cal O}(\alpha_S)$.  Because the
dominant gluon momenta are hard, we make the subsequent approximations:
i) $\gamma_T$ is independent of the momentum and is given by [\ref{Pis}]
\begin{eqnarray}
\gamma_T &=& {g^2NT\over 8\pi}
             \left( \ln{\omega_0^2\over m_{mag}^2 + 2m_{mag}\gamma_T}
             + 1.09681\ldots \right) \qquad \mbox{when} \ m_{mag}\gg\gamma_T,
\nonumber\\
         &=& {\alpha_S NT\over 2} \ln{1\over \alpha_S}
             + {\cal O}(\alpha_S T)
;\end{eqnarray}
ii) as we are only interested in the
leading order, we take $\mbox{Re}\Pi_T=3\omega_0^2/2$, as changing the
$\delta$-function $\delta(Q^2-\mbox{Re}\Pi_T)$ by a Lorentzian with width
$\gamma_T$ changes $\mbox{Re}\Pi_T=3\omega_0^2/2$ only by ${\cal O}(g^4T^2)$
corrections.
The gluon damping rate depends on the magnetic mass which is,
for dimensional reasons, a number times $g^2T$ [\ref{GPY}]. Fortunately,
this uncertainty enters only at the logarithmic level.

When $m=0$, we evaluate the integral over $q$ analytically, obtaining a
simple result:
\begin{equation}
R_T^{\mbox{q}} = \frac {2g^2 \gamma_T} {9\pi^4} \,
\int_0^\infty d\omega \, \omega^2 \, n_B (\omega) \, \left\{ \ln
\frac {64 \omega^4} {9\omega_0^4  + 16 \gamma_T^2 \omega^2}
+ \frac {3 \omega_0^2} {2 \gamma_T \omega} \left( \arctan \frac
{3 \omega_0^2} {4 \gamma_T \omega} + \arctan \frac {2\omega} {\gamma_T}
\right) - 4 \right\},
\end{equation}
where terms of order higher than $g^4$ have been dropped. Notice that
the limit $\gamma_T \to 0$ reproduces the result using a gluon
propagator without a finite damping rate.

The final result is
\begin{equation}
R_T^{\mbox{q}}
        = {4\zeta(3) \over 3\pi^3} \alpha_S^2
         \left( \ln{1\over \alpha_S} \right)^2 T^4
        + {\cal O} \left(\alpha_S^2 \ln{1\over \alpha_S} T^4\right).
\label{rtq}
\end{equation}
It is in fact the $Q^2$ dependence of the rate that makes it so large.
Should we use a bare gluon propagator, $\delta(Q^2)$, the rate would
identically vanish.  With just the hard thermal mass,
$\delta(Q^2-3\omega_0^2/2)$, the rate is of order $g^4T^4$ with no
logarithmic dependence. Taking into account the anomalously large
damping rate $\gamma_T$ shifts the gluon on-shellness $Q^2\sim \omega_0^2$
by a logarithmic correction. The additional logarithm has a kinematic
origin.

The processes that lead to our result are of the kind shown in Fig.~3.
The enhancement comes both from the soft space-like gluon exchange (a
Landau-damping term) and from the pole of the gluon propagator that is
almost on shell, hence the two logarithms. This diagram is not suppressed
kinematically.  Notice also that using a resummed propagator for a hard
line is not a new feature. For the calculation of the damping rate itself,
one needs to use the same propagator [\ref{Pis},\ref{DR}].

It is easy to see that the longitudinal decay is subleading.  The only
difference (aside from the self-energies) between the longitudinal and
transverse modes is a factor of two, as there are two transverse modes
and only one longitudinal mode. At hard momentum, the longitudinal
thermal mass is exponentially suppressed, $\mbox{Re} \Pi_L=0$, so we have
in the case of zero bare quark mass
\begin{equation}
R_L^{\mbox{q}} = \frac {g^2 \gamma_L} {9\pi^4} \,
\int_0^\infty d\omega \, \omega^2 \, n_B (\omega) \, \left\{ \ln
\frac {2\omega} {\gamma_L} - 4 \right\}.
\end{equation}
However, the longitudinal gluon damping rate is of order $g^2 T$ with no
log.  The quark production due to longitudinal gluon decay is therefore
of the same order as the neglected terms in eq.~(\ref{rtq}).

\bigskip
The discussion for massive quarks is the same through eq.~(\ref{elm}).
When $m\gg\omega_0$, the rate behaves perturbatively as $g^2 \gamma_T
\propto g^4\ln{1/g}$, and is therefore larger than other contributions.
We show in Fig.~4 the production rate as a function of $m/T$ for GP
($N_f=0$); the curves are very slightly altered for $N_f=2$.

We stress again that our result is the leading order perturbatively (when
$g\to 0$), containing all relevant contributions to order $\alpha_S^2
(\ln 1/\alpha_S)^2$.
The fact that the quark gluon plasma is not really a system of weak
coupling is a different problem, and should in principle be solved by
adding higher order contributions to our result.
One should work in this perturbative picture to make a reliable prediction.

In any case, there seems to be a huge uncertainty in the quark production
rate when considering large values of $g$ ($g>1$). As the rate is linearly
dependent on the damping rate, which itself depends on the value of the
magnetic mass, which is unknown, it is difficult at present to evaluate the
effect of higher order terms that are important for large $g$.
 We believe that
the rate can easily vary by an order of magnitude. One should bear this in
mind when making predictions for ultrarelativistic heavy ion collisions.
The disadvantage of this uncertainty can be turned into a gain, if one is
able to measure the thermal charm or bottom quark production. This would
give a direct measurement of the gluon damping rate, and hence of the
magnetic mass.

Having computed the quark production rate, we also discuss the
quark equilibration time given by $\tau_{\mbox{q}} = N_{eq}/ R^{\mbox{q}}_T$,
where $N_{eq}$ is the quark density at equilibrium.  In Fig.~5 we plot
$\tau_{\mbox{q}}$ as a function of $m/T$.  In calculating $N_{eq}$, we
take into account the complete mass (bare plus thermal corrections) at hard
momentum,
\begin{eqnarray}
m_\beta^2 &=& m^2 + {g^2 T^2\over 3} \qquad \mbox{when} \quad m \ll T,
\nonumber\\
m_\beta^2 &=& m^2 + {2 g^2 T^2\over 9}  \qquad \mbox{when} \quad m \gg T
.\end{eqnarray}
We cannot rigorously calculate the thermal mass between these limits, so
we use the ansatz
\begin{equation}
m_\beta^2 = m^2 + {(2+e^{-m/T}) g^2 T^2\over 9}.
\end{equation}
For vanishing quark mass, we obtain
\begin{equation}
\tau_{\mbox{q}} = {27\pi\over 16 \alpha_S^2 (\ln{1/\alpha_S})^2}
                  T^{-1}
.\end{equation}

For comparison, we also give the gluon equilibration time.  By detailed
balance, the thermal gluon formation rate must be equal to the decay rate
(twice the damping rate), so that
\begin{equation}
 \tau_{\mbox{g}} = {1\over 3 \alpha_S\ln{1/\alpha_S}} T^{-1}.
\end{equation}
We see that gluons equilibrate faster than even the lightest quarks.
This stems from the fact that quarks are always produced by pairs, which is
a relatively slow process, while gluons can be copiously emitted
by radiative processes.

Again, our results strongly depend on the value of the coupling constant.
{}From the numbers that we obtain, it seems clear that, in ultrarelativistic
heavy ion collisions, only the light quarks ($u$, $d$, and $s$) may have time
to come to chemical equilibrium, while the heavy quarks will not.  Even
taking $g=3$, $\tau_{\mbox{q}}>10 \, T^{-1}$ for massless quarks, so chemical
equilibration will proceed very slowly.

In conclusion, we have calculated the quark production rate from pure glue and
quark-gluon plasma in equilibrium. The leading contribution comes from the
decay of transverse gluons, which are mainly characterized by an anomalously
large damping rate $\gamma_T$.  The analogy of this process with the plasmon
decay into $\nu\bar\nu$ is very close but misleading. In the latter case, the
photon damping rate is smaller, $\gamma_{\gamma} = {\cal O}(e^3T\ln(1/e))$,
so the
plasmon is really described by a quasi-particle with an effective thermal mass.
 In a QCD plasma, this is clearly not the case, and one must be very careful in
computing the thermal rates by incorporating the damping rate inside the gluon
propagator.  Our result is another example of a physical quantity that is
sensitive to the non-perturbative magnetic mass scale [\ref{Pis}].

Although our calculation is perturbatively correct, there are large
uncertainties when making predictions for ultra-relativistic heavy ion
collisions for the following reasons:  i) the rate is linearly dependent on
 $\gamma_T$, which depends on the unknown magnetic mass, and ii) the
quark gluon plasma, which could be formed in such collisions is not really a
system of weak coupling, so higher order terms should be calculated.


\bigskip
This material is based upon work supported by the North Atlantic Treaty
 Organization under a grant awarded in 1991.


\ulsect{References}

\begin{list}{\arabic{enumi}.\hfill}{\setlength{\topsep}{0pt}
\setlength{\partopsep}{0pt} \setlength{\itemsep}{0pt}
\setlength{\parsep}{0pt} \setlength{\leftmargin}{\labelwidth}
\setlength{\rightmargin}{0pt} \setlength{\listparindent}{0pt}
\setlength{\itemindent}{0pt} \setlength{\labelsep}{0pt}
\usecounter{enumi}}

\item K. Geiger, Phys.\ Rev. {\bf D46} (1992) 4965. \label{rg}

\item E. Shuryak, Phys. Rev. Lett. {\bf 68} (1992) 3270. \label{rs}

\item J. Rafelski and B. M\"uller, Phys.\ Rev.\ Lett.\
{\bf 48} (1982) 1066. \label{rstrange}

\item A. Shor, Phys. Lett. {\bf B215} (1988) 375. \label{rcharm}

\item V.~V.~Klimov, Sov.\ J. Nucl.\ Phys.\ {\bf 33} (1981) 934;
      H.~A.~Weldon, Phys.\ Rev.\ {\bf D26} (1982) 1394;
      O.~Kalashnikov, Fortschr.\ Phys.\ {\bf 32} (1984) 525;
      R.~D.~Pisarski, Physica {\bf A158} (1989) 146.  \label{Plasmon}

\item E.~Braaten, Phys.\ Rev.\ Lett.\ {\bf 66} (1991) 1655. \label{Bra}

\item E.~Braaten and R.~D.~Pisarski, Nucl.\ Phys.\ {\bf B337} (1990) 569
      and {\bf B339} (1990) 310.\label{BP}

\item H.~A.~Weldon, Phys.\ Rev.\ {\bf D28} (1983) 2007. \label{Wel}

\item R.~L.~Kobes and G.~W.~Semenoff, Nucl.\ Phys.\ {\bf B260} (1985) 714
      and {\bf B272} (1986) 329;
      N.~Ashida, H.~Nakkagawa, A.~Ni\'egawa and H.~Yokota, Ann.\ Phys.\
      (NY) {\bf 215} (1992) 315. \label{KS}

\item J.~Kapusta, P.~Lichard and D.~Seibert, Phys.\ Rev.\ {\bf D44} (1991)
2744;
      R.~Baier, H.~Nakkagawa, A.~Ni\'egawa and K.~Redlich,
      Z. Phys.\ {\bf C53} (1992) 433. \label{Photon}

\item B.~L.~Combridge, Nucl.\ Phys.\ {\bf B151} (1979) 429. \label{Com}

\item R.~Pisarski, Phys.\ Rev.\ Lett.\ {\bf 63} (1989) 1129;
      R.~Pisarski, BNL preprint BNL-P-1/92. \label{Pis}

\item D.~J.~Gross, R.~D.~Pisarski and L.~G.~Yaffe, Rev. Mod. Phys.
      {\bf 53} (1981) 43. \label{GPY}

\item V.~V.~Lebedev and A.~V.~Smilga, Ann.\ Phys.\ {\bf 202} (1990) 229;
       T.~Altherr, E.~Petitgirard and T.~del Rio Gaztelurrutia, Phys.\ Rev.\
      {\bf D47} (1993) 703. \label{DR}
\end{list}


\ulsect{Figure captions}

\begin{list}{\arabic{enumi}.\hfill}{\setlength{\topsep}{0pt}
\setlength{\partopsep}{0pt} \setlength{\itemsep}{0pt}
\setlength{\parsep}{0pt} \setlength{\leftmargin}{\labelwidth}
\setlength{\rightmargin}{0pt} \setlength{\listparindent}{0pt}
\setlength{\itemindent}{0pt} \setlength{\labelsep}{0pt}
\usecounter{enumi}}

\item The lowest-order ``cut'' diagram for thermal quark
production.

\item The na\"\i ve lowest order Feynman diagrams for quark-antiquark
production.

\item A Feynman diagram contributing to leading order for thermal quark
production.

\item The quark production rate, $R^{\mbox{q}}_T$, in a GP ($N_f=0$) as a
function of $m/T$.

\item The quark chemical equilibration time, $\tau_{\mbox{q}}$, in a GP as a
function of $m/T$.

\end{list}

\vfill \eject

\end{document}